\documentclass[a4paper,fleqn,usenatbib,useAMS]{mnras}

\usepackage{graphicx}	% Including figure files
\usepackage{amsmath}	% Advanced maths commands
\usepackage{amssymb}	% Extra maths symbols
\usepackage{multicol}        % Multi-column entries in tables
\usepackage{bm}		% Bold maths symbols, including upright Greek
\usepackage{pdflscape}	% Landscape pages
\usepackage{subfig}

\def\eg{{\it e.g.,} }
\def\etal{{\it et al.}}
\def\ie{{\it i.e.,} }

\def\ha{H{$\alpha$}}
\def\hb{H{$\beta$}}
\def\hi{H\,{\sc i}}
\def\hii{H\,{\sc ii}}
\def\heii{He\,{\sc ii}}
\def\niii{N\,{\sc iii}}
\def\civ{C\,{\sc iv}}
\def\nv{N\,{\sc v}}

\def\ovi{O\,{\sc vi}}
\def\kms{km\hspace{1pt}s$^{-1}$}

\def\lsim{\lower.5ex\hbox{$\; \buildrel < \over \sim \;$}}
\def\gsim{\lower.5ex\hbox{$\; \buildrel > \over \sim \;$}}
\title[{\hi} Raman profiles of AG Peg]{The {\ha} and {\hb} Raman-scattering line profiles of the symbiotic star AG Peg}

\author[Lee and Hyung]{Seong-Jae Lee and Siek Hyung\thanks{Contact e-mail: \href{mailto:hyung@chungbuk.ac.kr}hyung@chungbuk.ac.kr} \\
School of Science Education, Chungbuk National University,
Chungbuk 28644, S. Korea}

\date{Accepted 2017 December 31. Received 2017 December 25; in original form 2017 September 26}

\pubyear{2017}

\begin{document}
\label{firstpage}
\pagerange{\pageref{firstpage}--\pageref{lastpage}}
%\frac{}{}
\maketitle

\begin{abstract}
\noindent
The  {\ha} and {\hb} line profiles of the symbiotic star AG Peg observed on 1998 September (phase $\phi$ = 10.24), display the top narrow double Gaussian
and bottom broad (FWHM of 200 -- 400 {\kms}) components.
The photo-ionization model indicated that the ionized zone, responsible for the hydrogen Balmer and Lyman lines, is radiation-bounded, with a hydrogen gas number density of $n_{\rm H} \sim10^{9.85}$ cm$^{-3}$ and a gas temperature  of T$_e$ = 12\,000 -- 15\,000~K.
We carried out Monte Carlo simulations to fit the Raman scattering broad wings, assuming that the
hydrogen Ly$\beta$ and Ly$\gamma$ lines emitted within the radiation-bounded {\hii} zone around a white dwarf have the same double Gaussian line profile shape as the hydrogen Balmer lines.
The simulation shows that the scattering {\hi} zones are attached to (or located just outside of) the inner {\hii} shells.
The best fit to the observed broad {\hi} line profiles indicates that  the column density of the scattering neutral zone is  $N_{\rm H} \simeq $ 3 -- 5 $\times$ 10$^{19}$ cm$^{-2}$.
We examined whether the geometrical structure responsible for the observed {\ha} and {\hb} line profiles is a bipolar conical shell structure, consisting of the radiation-bounded ionized zone and the outer material bounded neutral zone.  The expanding bipolar structure might be two opposite regions of the common envelope or the outer shell of the Roche lobe around the hot white dwarf, formed through the mass inflows from the giant star and pushed out by the fast winds from the hot white dwarf.

\end{abstract}

\begin{keywords}
H II regions -- binaries: symbiotic -- winds, outflows -- individual (AG Peg)
\end{keywords}

\section{INTRODUCTION}

Symbiotic stars are the most interesting because some systems host
the most massive white dwarf (WD), like SN Ia progenitors \citep{mun94, bof94}.
The slow symbiotic star AG Peg is a binary system that consists of a massive M3 III red giant star (GS) and a less massive hot WD surrounded by nebulous gas. The orbital period of AG Peg is known as  816.5 days.
The recent major outburst occurred in 2015, which is the second major one since the first major nova outburst in 1850 (see \citealt{ken01}). AG Peg appears to be  in a  continuous nuclear burning after the 1850 outburst, although it did not reach the temperature to be detected as a super-soft ($\leq $0.4 keV) X-ray source. The duration of the second outburst was relatively short.

The symbiotic star becomes an ideal object to study the mass loss rate from the GS and inflow into the WD, forming an accretion disk. High-speed outflow activities of massive dense winds from the GS and the signs of their interaction with hot WD zone gases were found. Various X-ray observations and the on-board Swift satellite also detected variability on a time-scale of days
\citep{all81, mur95, mur97, lun13, zhe16}.
Such a variability is likely to be caused by the shock interaction between the mass loss from the GS and the earlier ejected gas. \citet{mur97} detected 16 symbiotic stars and suggested a classification scheme based on the hardness of the spectra. According to their classification, AG Peg is a $\beta$-type with X-ray spectra with peak at 0.8 keV that might originate in a region where the winds from the two stars collide.

The presence of broad or Raman scattering features in the {\ha} and {\hb} spectrum
must be correlated with a source of high temperature photons, for which nuclear burning on the WD surface is the best candidate. According to \citet{con97}, the broad lines are emitted from swiftly moving photoionized gas surrounding the WD,
while narrow lines are emitted by shocked gas near the GS. The alternative kinematic interpretation of high-speed outflow for the broad line components
might be the line broadening effect due to Raman scattering process.
Modeling the wind structure and fitting the observed line profiles
would help us to clarify the ambiguity of the geometrical structure
and location of the emission zone(s) \citep{mer51, mer59, bel85, ken87, fek00, kim08, mun13}.

We analyzed the high dispersion line profiles of {\ha} and {\hb} of AG Peg, secured with the Hamilton Echelle Spectrograph (HES) at Lick observatory, which display two narrow and one broad wing components in the {\ha} and {\hb} lines. These broad wing features were investigated for other symbiotic stars or planetary nebulae in other studies \citep{lee00}. Monte Carlo simulations were done to fit the broad wing components which would give us the information of the location and structure of the {\hii} and scattering neutral hydrogen ({\hi}) zones.

In Section 2, we present the deconvolved {\ha} and {\hb} line profiles from the Lick Observatory Hamilton Echelle Spectrograph (HES) spectral data. In Section 3, we present the theoretical Lyman and Balmer line intensity ratios based on the predicted by the photoionization model for the observed line intensities. In Sections 4, we briefly describe Raman scattering process occurring in the {\hi} shell  responsible for the broad wing components along with explanation of Monte Carlo simulation procedure. Section 5 presents some of the predicted line profiles from the simulation and discusses the physical condition for the scattering zone. From the Raman scattering efficiency obtained through Monte Carlo simulation, one can also infer the ionization luminosity and temperature of the WD. Section 6 discusses possible geometrical structures responsible for the observed double Gaussian and broad wing profiles.
Some concluding remarks are given in Section 7.

\section{HES SPECTROSCOPIC DATA}

\begin{figure}
\includegraphics[width = \columnwidth,clip]{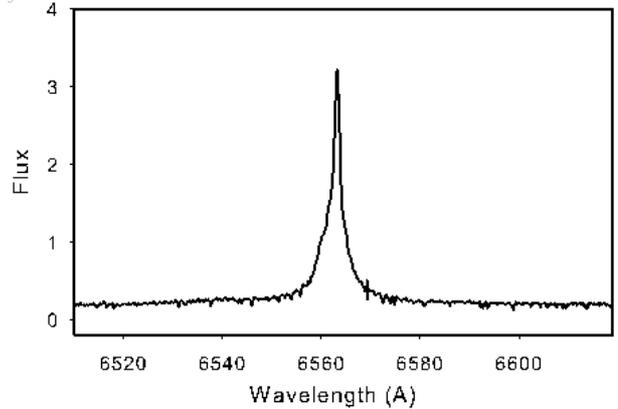}
\caption{Observed spectral scan showing the {\ha} emission line. Obs = 1998 ($\phi$ = 0.24). Exposure = 300 sec.  Flux unit: 10$^{-11}$ erg s$^{-1}$ cm$^{-2}$ \AA$^{-1}$.}
\label{fig1}
\end{figure}

The data used here are the spectra observed in 1998 September 17 by Aller and Hyung with the HES attached to the 3 m optical telescope at  Lick observatory. The employed slit entrance is 5$'' \times 1.5''$: with $1.5''$ (640micron) being the slit width.
%corresponds to the wavelength dispersion, \eg 0.2\AA\ /pixel at 5000\AA.
Both long and short exposures, 1800 and 300 seconds were necessary to have high signal-to-noise data and to avoid  saturation. Fig.~\ref{fig1} shows one spectral scan of AG Peg.

%** Figure 1 **

Based on the AG Peg's period, P = 816.5 day, the orbital phase ($\phi$) for the 1998 observation is obtained from the following
$$
%Max {\rm ($V$)} = JD 2,442,710.1 + 816.5  E,
Max {\it (V)} = JD 2,442,710.1 + 816.5  E,
$$
where Ephemeris $E$ defines the phase $\phi = 0.0 $  when $V$ reaches its maximum \citep{fer85, ibe96}.

We reduced the observed data by using IRAF (Image Reduction and Analysis Facilities) developed by NOAO (National Optical Astronomy Observatory). The  Lick HES high dispersion {\hi} line profiles are the stronger lines among various emission lines in the 3470 -- 9775\AA\, wavelengths which might be from the spatially unresolved compact zone  $<0.1''$. Detailed  description of observations is given by Kim \& Hyung (2008, KH08) and IRAF reduction procedures are found in \citet{hyu94}.  Table~1 lists the measured  {\ha} and {\hb} fluxes of the Lick observatory HES spectra. The observed absolute fluxes are F({\ha}) = 1.25 $\times 10^{-10}$  erg$^{-1}$ s$^{-1}$ cm$^{-2}$ [hereafter, 1.25(-10)] and F({\hb}) = 1.88(-11), respectively.

%** table 1**

\begin{table*}
\caption{Decomposed component fluxes  of the {\ha} and {\hb} lines and corresponding photon numbers.
1.03(-11) means 1.03 $\times 10^{-11}$ erg$^{-1}$ s$^{-1}$ cm$^{-2}$.  The flux values in brackets are interstellar extinction corrected, while the values in parentheses are the corresponding photon numbers per second.
% the blue+read flux errors are less than 2\%.
The {\hb} flux measurement was performed while avoiding {\niii} 4858 and {\heii} 4859 lines. The interstellar extinction corrected I({\ha})/I({\hb}) ratio for the (blue+red) flux (in the 4th column) is 2.87. The wide wing fractions for the {\ha} and {\hb} line profiles are 65.5\% and 20.3\%, respectively.
See the text and Fig.~\ref{fig2}. }
\vspace{-0.5cm}
\begin{tabular}{lcccccc}\\
\hline \hline

line & blue & red & blue+red & wing & total      & ratio \\
     &      &     &          &      & (blue+red+wing) & (blue+red): wing \\

\hline
{\ha} & 1.03$\pm$0.02(-11) & 3.01$\pm$0.02(-11)& 4.04(-11) & 7.43$\pm$0.07(-11) & 1.15(-10) & \\
        &                    &    & [4.30(-11)]  & [8.15(-11)]   & [1.25(-10)] & [34.5\% : { 65.5\%}] \\
        &                    &       & (14.2)  & (27.0)        &   &  \\

{\hb}  & 5.43$\pm$0.03(-12) & 8.27$\pm$0.04(-12) & 1.37(-11) & 3.49$\pm$0.11(-12) & 1.72(-11) & \\
        &            &                  &[1.50(-11) ]   & [3.82(-12)] & [1.88(-11)] & [79.7\% : { 20.3\%}] \\
        &                    &           & (3.68)  & (0.936)        &  &  \\

\hline

\end{tabular} \\
\label{tbl}
\end{table*}

The observed emission fluxes vary depending on the phase and they also show a long term variation. Kenyon {\it et al.} (1993, Table 5) list the measured {\hb} flux F({\hb}) =  3.82(-11) -- 11.26(-11) for the observation phases $\phi$ = 2.72 -- 7.12 (1981 -- 1991). Such a variation is likely to be brought on by the change of the WD luminosity and the physical state of the ionized gas zone. Analysis of the IUE spectral data secured from 1978 to 1995 by \citet{alt97} indicates that the WD varied its luminosity from 1850 L$_\odot$ to 430 L$_\odot$ but did not change its temperature much.

\begin{figure*}
\includegraphics[width = 11 cm]{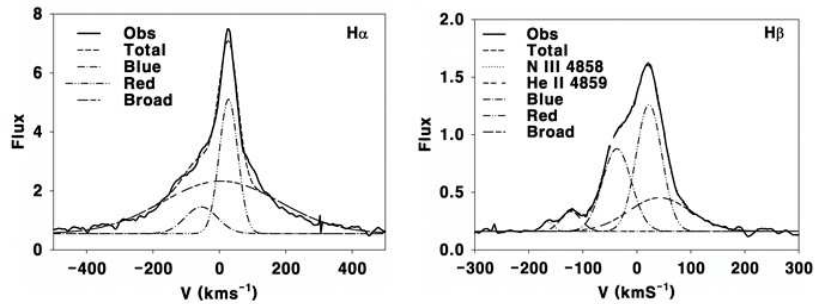}
\caption{{\ha} and {\hb} spectral line profiles. Observation year: 1998 ($\phi$ = 0.24). Exposures = 300 sec and 1800 sec. Flux: interstellar extinction corrected. Flux unit: 10$^{-13}$ erg s$^{-1}$ cm$^{-2}$ per radial velocity in {\kms}. The decomposed {\hb} profile also shows the other nearby {\niii} 4858 and
{\heii} 4859 lines.}
\label{fig2}
\end{figure*}

The {\hi} lines were deconvolved into three components with the help of  StarLink/Dipso developed by ESO (European Southern Observatory) and IDL (Interactive Data Language). Fig.~\ref{fig2} shows the decomposed profiles of the {\ha} and {\hb} line profiles with the Gaussian profiles. Note that {\niii} 4858 and {\heii} 4859 lines also exist at $-$165.3 and $-$123 {\kms} on the left side  of the {\hb} line. In decomposing the {\hb} profile with the Starlink/Dipso program, the {\niii}  and {\heii} components were carefully kept off (see Fig.~\ref{fig2}). The short 300 sec exposure proves to be useful for the analysis of strong lines in case when the strong lines suffer from the saturation in the long exposure. The short exposure data might be sometimes useful when one wants to monitor any variation feature or absorption features from the outer part, as ascertained in the IUE data \citep{eri04}. Our HES data do not show any absorption feature in spite of the comparatively high resolution capability of Lick HES, \ie $R$ $\sim$ 60\,000 (for a slit width of 1.2$''$). The comparison of the 300 sec and 1800 sec exposure fluxes in Table~1  shows agreement within 3.0\%, indicating the 30 min exposure is  not saturated (see Table~3). We used the 300 sec exposure  measurement  for both {\ha} and {\hb} in this work.

%** Figure 2 **

We derived the radial velocity of AG Peg based on the median wavelength point of the observed line fluxes, after correcting the Earth movement: \ie V$_{\rm r} = V_{\rm helio} = -$9.16 {\kms} (300 sec).
%Considering the flux ratio for the double Gaussian line profiles, we derived $V_{\rm r}$ = $-$9.09$\pm$0.14 {\kms} for the observed line profiles.
The derived radial velocity, $V_{\rm r} = -$9.09 {\kms}, has been
adjusted for the 1998 HES spectral data (phase $\phi = 10.24$, hereafter, 0.24)
and we define the zero (or the  assumed rest frame of AG Peg) in the horizontal axis of Fig.~2. The observed total {\hb} flux is the interstellar extinction corrected value, while the horizontal axis corresponds to the velocity ({\kms}) instead of wavelength (\AA), useful in deducing kinematic information from both profiles. The radial velocities found in the literature are from $-14.33$ to $-$25.00 {\kms}, \eg \citet{ike04}. We also derived the kinematically more useful radial velocity,  $V_{\rm r}$ = $-19.09\pm$0.14  {\kms}, from the middle of the double Gaussian line profiles.

{% when $-$17.09 used. {\ha}: 8.16,  {\hb}: 7.86   left sides of Lab-wavelength. shifted to the right.  therefore, $V_{\rm r}$ = $-$9.09 $\pm$error 0.14}

Both the {\ha} and {\hb} line profiles consist of three portions: top double Gaussian plus bottom broad components. The top double profiles consist of two well-defined blue- and red-shifted Gaussian type components, while the bottom portion shows a much broader single Gaussian wing. The {\ha} profile displays broad widths, indicating faster gas flow motion or complex physical condition of the line forming shell.
%As mentioned, we considered two cases for {\hb}  line profiles, \ie
% for the possibility of the presence of {\niii} and {\heii} lines.
%Interstellar extinction corrected flux values are also given in brackets.
The red components are 2.9 and 1.5 times more substantial than the blue component, for the {\ha} and {\hb} lines, respectively. The average ratio of the red to blue flux is about 2.2.

The interstellar extinction correction has been applied to the observed {\ha} and {\hb} fluxes with the interstellar extinction coefficient $C$ ($=$ log I({\hb})/F({\hb})) = 0.04 (or color excess E(B-V) = 0.027) adopted from KH08, where I({\hb}) and F({\hb}) are intrinsic and observed fluxes, respectively. The interstellar extinction corrected {\ha} and {\hb} values would be I({\hb}) = F({\hb}) $\times$ 10$^{C}$ and I({\ha}) = F({\ha}) $\times$  10$^{C(1-0.323)}$, respectively, taking on the wavelength extinction curve parameter f$_{\lambda}$ of \citet{sea79}. The corrected fluxes are given in brackets of Table~1 for the (blue+red) {\ha} and (blue+red)  {\hb} and the broad wing components, individually.

The interstellar extinction corrected ratio of the total flux (the 6th column in Table~1) is  I({\ha})/I({\hb}) = 1.25(-10)/1.88(-11) = 6.62, which is much higher than the recombination line theory, $\sim$3. Such large disagreements were also found in the earlier observations, \eg \citet{ken93}. The severe disagreement also exists in the {\ha} and {\hb} widths, implying complex kinematics of the emission zone. The wing flux fractions  for the {\ha}  and {\hb}  lines are 65.5\%  and 20.3\%, respectively (the 7th column in Table 1). Such a large discrepancy between the two implies that the broad components are not formed solely through a recombination mechanism. Meanwhile, the (blue+red) flux ratio is 2.87 (the 4th column in Table 1), close to the theoretical prediction, implying their formation
through the recombination mechanism.

A number of theoretical modeling studies have been performed to meet the observed line profiles for AG Peg. For example, Contini (1997) built a combined model consisting of three zones: (1) photo-ionized gas surrounding the WD for broad lines, (2)  shocked gas near the GS for narrow lines, (3) radiation-bounded region surrounding the WD for weak low-ionization UV lines and optical lines, while  \citet{eri04}  presented a more sophisticated model, which will be addressed in Section 4.1. Meanwhile, KH08 employed a single shell photoionization model to predict all the observed emission lines of the HES optical region spectral data secured during three different epochs. In this study, we considered the top double Gaussian profiles to be of nebular origin formed through the recombination mechanism, whereas the bottom broad wing profiles are formed through  Raman scattering process (not pure recombination lines).

Fig.~\ref{fig3} illustrates the face-on view diagram of AG Peg, which depicts the schematic representation of the relative spatial relations of the GS and the WD of AG Peg at phase $\phi$ = 0.24 for the 1998 observation (the observer at the bottom). It assumes a semi-detached binary system, establishing the mass outflow from the GS and the formation of the accretion disk model geometry around the WD.
Assuming the orbital axis is aligned, parallel  to the sky (\ie the inclination angle of the axis relative to the observer  is $i = 90^{\circ}$), the observer at the bottom  in Fig.~\ref{fig3} would be able to view both the GS and the WD along with the accretion disk. At the present $\phi$ = 0.24, we  must see both stars and the ionized nebular gas shell on the right side of the GS.

If the binary system is rotated to the other position at $\phi \sim$ 0.5, the observer would pick up the GS only, blocking the WD and the accretion. At $\phi \sim$ 0 (0.9 -- 0.1), the observer would pick up the bright WD + the accretion disk in front of the GS. The double Gaussian line profiles are present as well at phase $\phi \sim$ 0.56 and $\phi \sim$ 0.98 (KH08, unpublished), which means the inclination of the orbital axis, $i \neq 90^{\circ}$.

The detailed investigation on the dynamics or kinematic behavior of the gas shell responsible for the observed line profiles is beyond the scope of the present subject. Yet, one must invoke the approaching and receding zones of the nebular shell structure responsible for the double Gaussian line profiles.
We took one of two simple kinematic geometries (1) a rotating accretion disk around the WD and (2) a bipolar conic outflow structure, for the observed double peaks.
We will simply adopt the observed double Gaussian profile shape for the recombination lines and extend out the simulation of the Raman scattering process, implicitly adopting one of the two aforementioned structures.
%Later, we will give a discussion as to which geometry is better
% for the observed line profiles.

%** Figure 3 **

\begin{figure}
\includegraphics[width = \columnwidth,clip]{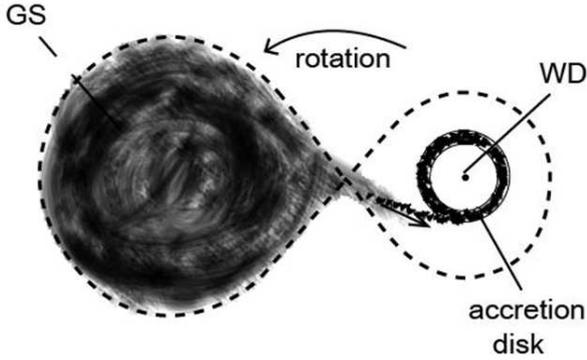}
\caption{Schematic diagram of the face-on view of the accretion disk
($i = 0^{\circ}$). The dashed line indicates the Roche lobe, taking on a semi-detached binary system. The real system could be a detached system depending on the GS status variation. The binary system and the accretion disk rotate in counter-clockwise direction. The observer is at the bottom. The left border of the accretion disk, closer to the GS, would appear to approach the observer.
%See the text for more details.
}
\label{fig3}
\end{figure}

\section{{\hi} RECOMBINATION LINES}
\subsection{Direct and scattered line components}

As seen in Fig.~\ref{fig2}, the double Gaussian line profiles appear in both the {\ha} and {\hb} lines.  The other Balmer line profiles of higher principal quantum numbers $n$ ($n \rightarrow$ 2; $n >$ 4), such as H$\gamma$ (4340.47), H$\delta$ (4101.74), and H$\epsilon$ (3970.07), also show both blue- and red-shift Gaussian components in our data. Here, the red components of these lines are likewise stronger than the blue ones, similarly as in the {\ha} and {\hb} lines.
Nonetheless, the broad wing profile components did not show up in these higher $n$ Balmer lines. \citet{ike04} observed the double Gaussian line profile spectra at phase, $\phi$ = 0.45. They understood that the GS in front of the emission zone blocked (or absorbed) the middle part of the wide emission lines, and as a result the lines became double peaks in shape. Nevertheless, the similar double Gaussian line profiles appeared in other orbital phases as in the present Lick HES data.

Considering the observed double Gaussian profiles and the expected relative position of the ionized zone (and two stars) at $\phi =$ 0.24, we construe that the observed blue- and red-components of the line profiles are either from a rotating accretion disk shell or from a bipolar conic shell, not by the blocking by the GS. The nebular lines must be formed from ionized gases in the zone close to the hot WD, mostly through the recombination process.
The line width seen in the top double Gaussian components might be associated to the thickness of the rotating accretion disk or the expanding bipolar conic shells.
The assumed geometry must accommodate the observed double Gaussian profiles in other phases, as mentioned.
%, \eg the high dispersion HES spectra of  2001 ($\phi \sim$ 0.56) and 2002 ($\phi \sim$
% 0.98) observations (KH08).

Our objective is to see whether one can simulate the broad line width seen at the bottom component through the Raman scattering process occurring in the nearby {\hi} zone, instead of assuming other faster moving structures.
Hence, we grouped the observed line profiles into two parts: (a) the blue and red profiles together as one part, formed through the recombination process and (b) the bottom broad profile as a second character, formed through the scattering process.
The similar {\ha}/{\hb} flux ratios of the double Gaussian profiles present in other phases made us believe that the emission zone is ionized by a relatively closer WD, not in other zone such as the colliding zones between the GS and the WD.
% or the extended atmosphere of the GS.

If the aforementioned shell is filled with a relatively high number density gas, $n_{\rm H} \sim$ 10$^9$ -- 10$^{10}$ cm$^{-3}$,  the central star might be not able to ionize the entire gas cloud, and as a result it might be radiation-bounded at least in a certain direction  \citep{con97}. Hence, we considered the radiation-bounded  shell geometry for the {\hii} zone in the simulation (referring to the P-I investigation result by KH08). The adjacent {\hi} zone outside the {\hii} shell might  be occupied by similar  high density gas, which becomes a favorable zone for Raman scattering, eventually forming the broad wing component.

\subsection{Balmer and  Lyman photons in the {\hii} shell}

One needs to know the number of UV photons entering into the {\hi} zone, which will experience the Raman scattering process. Two possibilities for the UV photon source are usually considered: (1) the spectral energy distribution (SED) of the WD and (2) strong hydrogen Lyman lines emitted in the {\hii} zone. The UV continuum flux levels of the SED around the Ly$\beta$ and Ly$\gamma$ wavelengths are too low for the Raman scattering process responsible for the broad {\ha} and {\hb} components (see \citealt{leh00}). Hence, we need to estimate the latter case, the number of Ly$\beta$ and Ly$\gamma$ photons formed in the ionized shell. However, one cannot directly observe the Lyman line intensities emitted in the ionized zone.

Although the hydrogen Lyman series lines are not easily observable, we can theoretically estimate these line fluxes by a recombination theory. The observed Balmer line information listed  in Table~1 will become a base in estimating the Lyman photon numbers formed in the ionized zone and we expect that the Lyman lines have  double Gaussian components similar to the Balmer lines since they are from the same zone with the same kinematics.

%, after  experiencing the scattering process?

%Using similar physical variables for shell gas densities and WD temperature as the one given in P-I models by KH08 (see their Table~5), we can calculate Hydrogen emission fluxes predicted by using CLOUDY (or recombination theory).

Table~\ref{tb2} lists the intensities for Ly$\alpha$, Ly$\beta$, Ly$\gamma$, and {\ha} relative to {\hb} for two electron temperatures in the ionized shell:  T$_{e}$ = 12\,000 and 15\,000~K.  Note the predicted {\ha}/{\hb} ratio which is close to the observed (blue+red) Balmer ratio I({\ha})/I({\hb}) = 2.87 in Table~1. The intensities of the Lyman series lines, especially the higher principle quantum numbers  $n$ ($n \rightarrow$ 1; $n >$ 3),  are very sensitive to the gas number density of the ionized zone, quite different from Balmer line intensities (see \citealt{leh00}). Our calculation confirms that the hydrogen gas number density, $n_{\rm H}$ = $10^{9.85}$ cm$^{-3}$ as in KH08, gives the appropriate Lyman line ratios.

%** table 2 **

\begin{table}
\caption{Theoretical {\hi} flux ratios in the ionized shell.
Flux ratios are presented on the base of I({\hb}) = 1.
Relative photon number ratios are also given in brackets.
The observed (blue+red) flux ratio I(\ha)/I(\hb) = 2.87 (in Table 1) is close to
the theoretical flux ratios I(\ha)/I(\hb) = 2.92 and 2.65 for T$_e$ = 12\,000~K and 15\,000~K, respectively, but the observed total or bottom wing fluxes disagree largely. See the text.}
\vspace{-0.5cm}
\begin{tabular}{ccccc}\\
\hline \hline

T$_{e}$ (K) & I$_{Ly\alpha}$ & I$_{Ly\beta}$ & I$_{Ly\gamma}$ & I$_{H\alpha}$ \\

\hline
12\,000      &  69.0 [17.25]    &  8.34 [1.76]    &  3.16 [0.632]    &  2.92 [3.95]  \\
15\,000      &  47.6 [11.9]     &  7.58 [1.60]    &  3.19 [0.638]    &  2.65 [3.58]   \\
\hline

\end{tabular} \\
\label{tb2}
\end{table}

The Ly$\alpha$ photons emitted in the ionized zone would become a diffuse ionizing source for other elemental atoms and ions of low-lying ionization potentials, while Ly$\beta$ and Ly$\gamma$ photons emitted in the ionized zone may escape the {\hii} zone and enter into the neutral zone and suffer a number of scattering. Some fraction of them would finally transform into optical wavelength quanta through the Raman scattering process during their scattering process. Based on the P-I model result, we calculated Lyman and Balmer line intensity ratios. We will use the predicted line intensities for T$_{e}$ = 15\,000~K in the present investigation, which might be better physical conditions considering other high excitation lines, \eg {\heii},  {\ovi}, and UV lines observed in other works.

In the scattering process simulation, the information for emitting photon numbers is more useful than the flux value itself. Utilizing a single phone energy, \eg E({\ha}) = 1.89~eV (3.02 $\times$ 10$^{-12}$~erg), E({\hb}) = 2.55~eV (4.08 $\times$ 10$^{-12}$~erg), and other Lyman lines, we calculated the relative photon numbers [s$^{-1}$] formed in the {\hii} zone(s), which are given in brackets of Table~\ref{tb2}.

\section{RAMAN SCATTERING SIMULATION}

The nebular gas in symbiotic stars is known to be supplied by the GS wind,
which may form various structures, \ie an accretion disk and bright zones
between the two stars. Additional components of strong wind from the WD were also known to exist in AG Peg \citep{mur95}.

\subsection{Emission zones and double Gaussian lines}

Table~\ref{tb3} lists the measured full width half maximums (FWHMs) of the observed {\ha} and {\hb} lines, whose different line widths seem to hint that the emission zone for these lines consists of various or at least three kinematically different zones. The top two peaks have relatively narrow widths, while the bottom components show very wide line widths, up to $\sim$600 {\kms}. The different line widths and intensities of the top double peak profiles could be interpreted as being formed in physically and kinetically different zones.

%** table 3 **

\begin{table*}
\caption{FWHM and peak center of the line profiles.
FWHM: {\kms}.  The 30 min measurements are given in parentheses. The FWHMs of the {\ha} and {\hb} lines and their centers  are from the Gaussian profiles. $^*$: kinematic profile centers in parentheses (assuming the radial correction velocity with  $V_{\rm r}$ = $-19.09\pm$0.14).
Note that the center of the {\hb} broad wing differs from that of the {\ha}. See Fig.~\ref{fig2}.}
\vspace{-0.5cm}
\begin{tabular}{cccc}\\
\hline \hline

line & blue & red & wing  \\

\hline
{\ha}  &  114 at -54.7$\pm$3.3  &  66.2 at +27.8$\pm$0.4  &  415 at 5.1$\pm$2.3  \\
{\hb}  &  63.1 at -36.8$\pm$0.3  &  55.9 at +22.7$\pm$0.2  &  134.4 at 42.1$\pm$2.5  \\
       &  (61.8 at -38.2$\pm$0.3)  &  (58.9 at +21.9$\pm$0.2)  &  (130.0 at 57.3$\pm$2.3)  \\
mean$^*$  &   88.5 at -46.1 (-36.1)  &  61.1 at +25.1 (+35.1)  &   \\
\hline

\end{tabular} \\
\label{tb3}
\end{table*}

The analysis of the high dispersion spectra by \citet{nus95, ken93, eri04} discerns more than three different emission regions for AG Peg: (1) The WD wind zone, (\ie at a terminal velocity $\sim$1000 {\kms}) from the hot WD due to the slow nova eruption beginning in 1850; (2) two separated emission zones of relatively high density located in the GS wind of 30 to 100 {\kms} \citep{eri04}; and (3) the wind collision zone of lower density,  where the fast $\sim$700 {\kms} wind from the WD collides with the $\sim$60 {\kms} wind from the GS, which may make the composite line structure seen in $\sim$200 {\kms} {\civ} and {\nv} IUE doublets. X-ray emission harder than a few 100 eV was detected, which might be given off by a hot plasma with a temperature of a few million K shock-heated by colliding winds.
(4) The blue-shifted absorption zone from the wind regions as well as the surrounding nebula also  exists, affecting the UV lines and producing P Cygni profiles.

The broad wings seen in our Lick HES {\ha} and {\hb} line profiles correspond to the aforementioned first WD wind component, while the double Gaussian components correspond to the aforementioned second double emission zone components.
The deviation of the line widths between the {\ha} and {\hb} profiles might be caused by the rest zones, the third and fourth zones. For the relatively wider {\ha} profile, one might be able to extract these third and fourth components found in the Lick HES data. We did not scrutinize the third and the fourth components from both {\ha} and {\hb} profiles, since such additional components are not our main concern.

Other symbiotic stars such as RR Tel and V1016 Cyg, also exhibit double-peak profiles in {\ovi} 6825, 7082. \citet{lee99} and \citet{heo15} attributed the {\ovi} double-peak profiles to the kinematics of the accretion disk emission gas. The imbalance of the double-peak profiles might occur in a symbiotic binary system when the WD gravitation captures some fraction of the slow stellar wind from the GS \citep{dev09, mas98}. For their studies, \citet{lee99} and \citet{heo15} proposed that Raman scattering occurs in a high-density neutral gas zone near the GS.
%See the diagram where the observer at the bottom  in Fig.~\ref{fig3}.

%observed  at $\phi \sim$ 0.25
One possible alternative model geometries for the present double Gaussian line profiles is a rotating accretion disk around the WD. The other alternative model geometry considered in this study is an expanding bipolar conic shell structure, which also produces double Gaussian line features when the polar axis is slanted to the sky plane.

The {\hii} zone responsible for the hydrogen Balmer and Lyman line emissions is taken up to induce the same ionization balance, structure as the KH08 P-I spherical shell model. In the simulation, we will not explicitly specify any particular model geometry for the {\hii} zone responsible for the observed double Gaussian profiles.
We will simply adopt the same double Gaussian form for the hydrogen Balmer and Lyman lines in the calculation, assuming that the unspecified semi-spherical structure produced them.

\subsection{{\hii} shell dimension for
the observed {\hb} flux}

First, we may start with a simple spherical {\hii} shell, \ie KH08 P-I model, which might be improved at a final step by setting its size and shape suitable for the observed fluxes, \eg {\hb} flux. The neutral zone could be situated in the further out from the {\hii} shell, \eg the extended atmosphere of the GS or next to the outer boundary of the {\hii} shell! The simulation was executed in a spherical coordinate, treating the outer neutral shell and the inner {\hii} shell as one part of the spherical shell. The Lyman photons formed in the {\hii} shell are assumed to enter into the outer {\hi} spherical shell  on normal line directions.
The scattered photons will also come out in random directions from the surface of the {\hi} zone.
%, due to the unpredictable random scattering.

Table~\ref{tb4} presents models for AG Peg {\hii} zone and their corresponding parameters and predicted values. The assumed stellar temperature T$_{\rm eff}$ = 120\,000~K is the same in all models, only the assumed WD radii and corresponding luminosities are different. As remarked, the ionization structures in all models were chosen to be the same, \ie I({\ha})/I({\hb}) = 2.65 for T$_e$ $\sim$ 15\,000~K (see Table~~\ref{tb2}). Nonetheless, the nebular shell dimension, \ie the {\hii} shell radius and thickness (and emitted total photon numbers within the volume) is different. From the inner and outer radii information of the {\hii} shell of each model, we  can figure the total {\hb} flux or the photon number emitted in the {\hii} shell volume.

%** table 4 **

\begin{table*}
\caption{Spherical models for {\hii} zone.
3.16(13) means 3.16 $\times 10^{13}$ cm. All models have the same ionization balance structure, \ie the WD effective temperature T$_{\rm eff}$(K) = 120\,000 K and the
electron temperature of the {\hii} shell, T$_e$(K) =  15\,000~K.
% The assumed gas number density for the {\hii} shell in the P-I model
% is  $n_{\rm {\hii}}$  =  $10^{9.85}$cm$^{-3}$.
$^a$The {\hb} luminosity (erg s$^{-1}$). $^b$The predicted {\hb} flux at Earth [erg~s$^{-1}$~cm$^{-2}$], while the observed {\hb} flux F({\hb}) = 1.50($-$11) erg~s$^{-1}$~cm$^{-2}$.
See the text.}
\vspace{-0.5cm}
\begin{tabular}{llccc}\\
\hline \hline

& parameter & Model I & Model II & Model I-A \\

\hline
distance  & d (pc) &    1000     &  650  & 1000   \\
WD luminosity  &  L$_*$ (L$_{\odot}$)  &  300  & 400 &   60   \\
%     &  T$_{\rm eff}$ (K)        &     120\,000  &    120\,000 & & \\
\hline
% P-I    &  {\hii} T$_e$ (K)   &    15\,000 &    15\,000 &  & \\
{\hii} shell &      &       &  &    \\
inner radius  &    r$_{\rm i}$ [cm]  &  3.16(13)   & 3.65(13)  &  1.42(13)  \\
outer radius     &  r$_{\rm o}$ [cm]    &  3.178({13}) &  3.67(13) &  1.428(13)   \\
      &   r$_{\rm o}$ [AU]                     &   [2.1]  &  [2.45] &     [0.95]          \\
\hline
Luminosity  & {\hb}$^a$ &  8.91({33})   & 1.78(34) &    1.78(34) \\
Flux at Earth &  F({\hb})$_{prd}$$^b$  &  7.43($-$11)  & 2.34($-10$)  &  1.50($-11$)  \\
\hline

\end{tabular} \\
\label{tb4}
\end{table*}

The shell dimension of Model I is basically that of  KH08 P-I model. The WD luminosity is  L$_{\rm WD}$ = 300 L$_\odot$ and the  {\hii} shell radius is about 2.1 AU. The integrated {\hb} flux is 10$^{33.94}$ -- 10$^{33.95}$ erg s$^{-1}$.
This would give F({\hb})$_{prd}$ =  $7.26$ -- $7.43 \times 10^{-11}$ erg s$^{-1}$ cm$^{-2}$ at a distance of 1 kpc, about 5 times larger than the observed flux
at Earth,  $1.50 \times 10^{-11}$ erg s$^{-1}$ cm$^{-2}$.

\citet{vog94} derived the WD luminosity to L$_{\rm WD}$ = 400 -- 500 L$_{\odot}$ for  distance d = 650 pc determined based on the interstellar extinction E(B-V) = 0.09 (see also M{u}rset \etal 1991; Penston \& Allen 1985). Founded on these values,  L$_{\rm WD}$ = 400 L$_\odot$ and  d = 650 pc, the predicted flux at Earth (Model II) would be F({\hb})$_{prd}$ =  $2.34 \times 10^{-10}$ erg s$^{-1}$ cm$^{-2}$, which is also much larger than the observed value.
Both Model I and Model II are assumed to be a spherical shell.

By adapting the different WD luminosity (L$_*$ $\sim$ $\pi$ R$_*^2$ T$_{\rm eff}^4$) and the different shell dimension  accordingly (but keeping the ionization balance {the same} as in Model I), we can figure the physical parameters for the WD and shell to fit the observed flux. For example, to meet the observed {\hb} flux, the WD luminosity in Model I should decrease as L$_{\rm WD}$ = 300 $\times$ (1.5/7.43) $\sim$ 60 L$\odot$ (assuming a str\"{o}mgren {\hii} shell sphere around the WD) and the shell radius should be  r$_{\rm o}$ $\sim$ 0.9~AU  (from a relation r$_{\rm o}^2$ $\propto$ R$_*^2$) as in Model I-A. However, the adopted WD luminosity in Model I-A seems too much small, smaller than most other values indicated by other earlier works.

Since the ionization balance structures in Model I and Model II are fairly good in predicting the line intensities, we considered slightly different geometrical structures, \ie the aforementioned accretion disk shell and the polar cones,  which still maintain the same WD luminosity and the ionization structure, rather than adopting radically different geometrical parameters.

Table~\ref{tb5} lists the accretion disk and bipolar conic models refined by the additional parameters for Model I and Model II in Table~\ref{tb4}. All models would fit the observed {\hb} flux at Earth, by adopting the accretion disk or bipolar conic model geometries to confine the {\hii} emission zone in a small volume fraction. The Model I-B accretion disk   assumes an equatorial latitude angle of $\leq$11.5$^{\circ}$ (corresponding 20\% volume fraction), while  Model II-A requires a smaller (more reasonable) equatorial latitude angle, $\leq$3.7$^{\circ}$ (corresponding to a 6\% volume fraction). The alternative shell is a bipolar conic shell. The  opening angle, $OA$ = 74$^{\circ}$ in Model I (or Model I-C in Table~\ref{tb5}) would give the 20\% volume fraction of the spherical shell, while $OA$ = 40$^{\circ}$ in Model II (or Model II-B in Table~\ref{tb5}) would give the  6\% volume fraction necessary to meet the observed flux.
The Model II-A or Model II-B shell appears to be more appropriate than the Model I cases, considering the employed WD luminosity.

%** table 5 **

\begin{table*}
\caption{Accretion and bipolar models.
See Table~\ref{tb4} for additional parameters.
$^a$The latitude angle range from a torus shape accretion shell model and a bipolar cone model.
$^b$The corresponding volume fraction relative to the spherical shell.
$^c$Opening angle of the bipolar cones, to adjust the model volume fraction to fit the observed flux.
% The assumed gas number density for the {\hii} shell in the P-I model
% is  $n_{\rm {\hii}}$  =  $10^{9.85}$cm$^{-3}$.
%r1/A = 0.38 + 0.2 log(M1/m2) = 0.51*, 0.54* 3.3AU = 1.7 & 1.8 AU
$^d$The separation from the WD to L1 (for M3 III GS ($\sim$6~M$_\odot$) and WD ($\sim$1.4 M$_\odot$) to check the {\hii} accretion disk radius (in Model I-B and Model II-A). The radius of the shell is a distance from the WD, so the dimension (or diameter) of the rotating accretion disk (Fig.~\ref{fig4}) or the expanding bipolar cones (Fig.~\ref{fig7}) would be double the radius. }
\vspace{-0.5cm}
\begin{tabular}{llccc}\\
\hline \hline

parameter & Model I-B & Model II-A & Model I-C & Model II-B \\

\hline
geometry   &   accretion   & accretion &   bipolar   & bipolar\\
distance (pc)    &  1000 & 650  &  1000 & 650  \\
 shell radius (AU)      &  2.1 & 2.45 &  2.1 & 2.45    \\
 shell dimension (AU)      &  4.2 & 4.9 &  4.2 & 4.9    \\
 ~~~~~~~~~~---~~~~~~~~~ ({\arcsec})  &   0.004 & 0.008  &   0.004 & 0.008  \\
\hline
 disk (${|}{\rm \theta}{|}$)$^a$  &    $\leq11.5^{\rm \circ}$  &  $\leq3.7^{\rm \circ}$  & & \\
 fraction$^b$  &    20\% & 6\%   & & \\
\hline
bipolar cone (${|}{\rm \theta}{|}$)$^a$  &  & &  $\geq53.1^{\rm \circ}$  &  $\geq70.1^{\rm \circ}$ \\
  fraction$^b$  &  &  & 20\% & 6\%   \\
open angle$^c$  & &  &   [$74^{\rm \circ}$]  &  [$40^{\rm \circ}$] \\
\hline
M$_{\rm GS}$ : M$_{\rm WD}$ [M$_\odot$] &  6 : 1.4   &  6 : 1.4  &  & \\
semi-major (AU)  & 3.5 &3.5 & &  \\
L1-WD distance (AU)  & 1.6$^d$   & 1.6$^d$   & &  \\
\hline

\end{tabular} \\
\label{tb5}
\end{table*}

The 1.5 GHz and 5 GHz study by \citet{ken91} showed that AG Peg consists of the inner  $\sim$2$''$ and outer $\sim$20$''$ nebulae  and $\sim$1$'$ bipolar object.
The earlier model studies, \eg by \citet{con97}, assumed a nebula radius of  r$_n$ = 210 AU corresponding to the above inner $\sim$2$''$ radio image. Even so, the present work concerns a much smaller size inner {\hii} zone, very close to the WD.
The outer radii,  2.1 AU and 2.45 AU, of models in Table~5, are all radiation-bounded, whose  {\hii} shell diameters correspond to $\sim$0.0042$''$ and $\sim$0.0082$''$, respectively, at Earth.

\subsection{Monte Carlo simulation}

As remarked, the  double Gaussian line profiles in Fig.~\ref{fig2} are of pure nebular origin, while the bottom broad wings are of hydrogen Lyman photons formed in the {\hi} zone through the Raman scattering process. In this subdivision, we explain how Monte Carlo simulation was done in the Raman scattering process in the {\hi} zone to fit the observed broad wing components seen in {\ha} and {\hb} line profiles.

Raman scattering by atomic hydrogen was first investigated by \citet{sch89} for the broad features around 6825\AA\, and 7082\AA. When far UV photons at $\lambda_i$ shorter than Ly$\alpha$ are Raman scattered by neutral hydrogen atoms in an initial ground 1s state, the hydrogen atoms will be excited to upper layers {(n = 3 or 4)} and some fraction of them will subsequently de-excite to the 2s or 3s state, emitting optical wavelength photons at $\lambda_f$.
{For instance, a Ly$\gamma$ photon is transformed either into a P$\alpha$ photon plus an {\ha} (or into an {\hb} photon) plus a Ly$\alpha$ photon after a number of scatterings.
The absorption leading to excitation of the n = 5 or higher levels is ignored because of the very low population.}
The $\lambda_f$ of the outgoing photon is given by
$$
hc/\lambda_{f} = hc/\lambda_{i} - hc/\lambda_{\alpha},
$$

\noindent where $\lambda_{\alpha}$ is the scattered Ly$\alpha$ wavelength. This  leads to the following

$$
\frac{\triangle\lambda_f}{\lambda_f}
=
\frac{\lambda_f}{\lambda_i}\frac{\triangle\lambda_i}{\lambda_i},
$$

\noindent which implies the Raman scattered lines at $\lambda_f$ have a broad wing broadened by a factor of ${\lambda_f}/{\lambda_i}$. The relationship explains the Raman scattering broad wings around {\ha} and {\hb} in Fig.~2, respectively \citep{cha15, lee03, leh00, lee97}.

We presumed that the {\hii} shell emits from the center radially outward the Ly$\beta$ and Ly$\gamma$ (and {\ha} and {\hb}) lines in the shape of a double Gaussian profile. In the simulation, we use the information of photon numbers generated in the {\hii} zone. The total photon numbers of the {\hb} line and other Ly$\beta$, Ly$\gamma$, and {\ha} lines within the disk or bipolar cones are already estimated by the P-I model (see  Table 2). Now, we require specifications of their double Gaussian distribution form in the wavelength (or velocity) domain.

The {\ha} double Gaussian profile centers in Table~\ref{tb3} are at $-$54.7$\pm$3.3 and +27.8$\pm$0.4 {\kms}, while the {\hb} profile centers are at $-$37.5$\pm$0.3 and +22.3$\pm$0.2 {\kms} (the mean of the 5 min and 30 min).
The centers of the observed line profiles are at $-46.1$ {\kms} [(-54.7$-37.5$)/2] and 25.1 {\kms} [(27.8+22.3)/2], respectively. If we take the radial velocity, $V_{\rm r} $ = $-19.09\pm$0.14 instead of $-9.09\pm$0.14 {\kms}, the centers of the double Gaussian line profiles will be in kinematic symmetrical positions, $-$36.1 and +35.1 {\kms}, respectively.

In the simulation, the hydrogen Ly$\beta$ (or Ly$\gamma$) photons emitted in the {\hii} zone are assumed to be in the double Gaussian distribution form with two peaks at $v_{b}$ and $v_{r}$ = $-$35 and 35 (or at $-45$ and +25) {\kms} and with the FWHM 65 {\kms}, slightly different but similar to the observed mean values in Table~3. The hydrogen Balmer lines indicate that the flux of the red component is approximately 2.2 times stronger than the blue component (the average value from the {\ha} and {\hb} lines in Table~1). Based on the observed hydrogen {\ha} and {\hb} double Gaussian line profiles, we assumed the double Gaussian distribution form, as $N_{\rm Gaussian} $ = $N_{\rm total}$ $[$(1/3.2) $f_b$ +  (2.2/3.2) $f_r$$]$ for both Ly$\beta$ and Ly$\gamma$ lines (using the Lyman flux data information given in Table~2).
Here, the Gaussian distribution functions $f_b$ and $f_r$ for the blue and red Gaussian profiles are
\begin{eqnarray*}
f_{b} & = & 1/(\sqrt{2 \pi} \sigma_G)~exp[-\frac{(x - v_{b})^2}{2~\sigma^2}]
\end{eqnarray*}
and
\begin{eqnarray*}
f_{r} & = & 1/(\sqrt{2 \pi} \sigma_G)~exp[-\frac{(x - v_{r})^2}{2~\sigma^2}],
\end{eqnarray*}
where $\sigma_G$ is the Gaussian RMS width  (the same for both); the peak centers $v_{b}$ and $v_{r}$ are at $-45$ and +25 (or $-$35 and +35) {\kms} as mentioned above, respectively; and  $1/(\sqrt{2 \pi}\sigma_G)$ is the peak height. All the FWHMs (= 2$\sqrt{2~ln 2}~\sigma_G$) of the Lyman and  Balmer lines are taken to have the same 65 {\kms} (close to the mean values in Table~\ref{tb3}). The total photon numbers, $N$(Ly$\beta$) and $N$(Ly$\gamma$), were calculated based on their ratios to the {\hb} photon number given in Table~\ref{tb2}. With the adopted double Gaussian line emission distribution, we will no longer have concern about the asymmetry or whether the emissions are from the rotation of the accretion disk or from a bipolar conic shell.

%The hydrogen Lyman photons formed in the {\hii} zone would enter into
% the outer neutral hydrogen shell and experience the scattering.
Since the non-elastic cross-section of Ly$\beta$ or Ly$\gamma$ by {\hi} atoms is smaller than the Thompson cross-section, the Raman scattering line requires not only the aforementioned emission zone but also the somewhat thick {\hi}  zone.
As long as the scattering {\hi} zone is placed near by the {\hii} shell, it does not matter whether the scattering zone is attached or detached from it. Hence, the  inside radius of the neutral {\hi} shell, r$_{\rm i}$(H {\sc i}), may coincide with the radiation-bounded outer radius, r$_{\rm o}$(H {\sc ii}), of the {\hii} shell. The thickness of the neutral shell or the outer boundary r$_{\rm o}$(H {\sc i}) is to be inferred from its column density from Raman scattering simulations.

The shell thickness of the neutral zone, $\triangle$D, defined by the inner and outer radii, r$_{i}$({\hi}) and r$_o$({\hi}), is {\it a priori} information in the simulation. The gas density of the neutral zone of the disk responsible for the scattering process would be different from that of the {\hii} zone, but we do not need to know the exact value of gas density and shell thickness of the {\hi} shell, simultaneously. Instead, we will estimate the {\hi} column density $N_{\rm HI}$ (cm$^{-2}$) $\equiv$ {$\triangle$}D (cm) $\times$ $n_{\rm HI}$ (cm$^{-3}$) in the simulation until the prediction fits the observed broad line profiles.

We also re-scale the geometry of the neutral hydrogen shell {in terms of the total optical depths  $\tau_{\rm D}$ = $\sigma$($\lambda$)$\times n_{\rm HI}\times\triangle {\rm D}$  and   $\tau_{\rm H} = \sigma$($\lambda$)$\times n_{\rm HI}\times\triangle \rm H$ for both the shell thickness $\triangle {\rm D}$ and the shell height  (or arc length)} $\triangle$H. Here, the $\triangle$H is  actually not a vertical height as in cylindrical torus shell, but it fits to the curved latitude distance ($\triangle$H $\sim$ r$\times \theta$) along the latitude in a spherical coordinate (see Fig.~\ref{fig4} and Fig.~\ref{fig7}). We do not have to consider this $\triangle$H direction since most photons cannot escape through this  $\theta$ (or $\varphi$) direction in a torus type accretion disk  or in a bipolar conic shell, though!

We utilized the most recent calculation for the total cross sections and branching ratios around Ly$\beta$ and Ly$\gamma$ by \citet{cha15} (see their Fig.~1 and Table~2). By dividing  the $\lambda = $ 1011.83 -- 1035.69\AA\, interval around Ly-$\beta$ into about 3000 $\times$ $\triangle\lambda$ (= 0.008\AA) wavelength segments  around Ly$\beta$, we carried out the Monte Carlo simulation with the photons within each segment.
%(corresponding to $\triangle \lambda$ = 0.05\AA\, around {\ha}),

After coming in into the {\hi} zone, {a Lyman photon (\ie absorption leading to excitation of the n = 3 or n =4 )} travels until it meets a {\hi} atom for the first time and then it experiences either Rayleigh or Raman scattering. To determine the first scattering site, Monte Carlo simulation employs an estimation of an optical depth $\tau$ by
$$
\tau  =  -ln~q
$$
where $q$ is a random number uniformly distributed in the interval [0,1]. The estimation of the optical paths, $\tau_{Ly\beta}$ or $\tau_{Ly\gamma}$, for the traveling distance  by Ly$\beta$ or Ly$\gamma$ photons and their  random scattering directions, are found in a spherical coordinate ($r$, $\theta$, $\varphi$). (1) If the Lyman photon transforms into the optical photon through the Raman scattering process, it can escape the neutral zone like other Balmer photons in the inner {\hii} zone: the relatively thin outer {\hi} shell is assumed to be transparent to the optical emission lines. (2) Meanwhile, if it experiences Rayleigh scattering, the Rayleigh scattering Lyman photon will go into another random direction until it meets another {\hi} atom. (3) If the Rayleigh scattered photon (still being a Lyman photon) is nonetheless within the neutral zone, it will travel until it meets another {\hi} atom and experiences another scattering, either Rayleigh scattering or Raman scattering. The above procedure will be repeated until all the Lyman photons transform into Raman scattering optical lines or escape the scattering zone.

The height (or arc length) to thickness ratio of the {\hi} shell in either an accretion disk
or a bipolar conic shell is
$A$ = $\triangle$H/$\triangle {\rm D}$ $\simeq$  (r$_i \times \theta$)/$\triangle {\rm D}$ =
(3.17 $\times 10^{13} \times$ $11.5^{\circ}$)/(1.8 $\times 10^{11}$) $\sim$ 70 for the latitude ranges from $\theta$ = 0$^{\circ}$ to $\pm$11.5$^{\circ}$ in Model I-B (see Fig.~\ref{fig4}).
Model II-A will also have a similarly high ratio, $A$ $\sim$ 25.
When the `$A$' value exceeds 4, the escape probability {depends on the shell thickness only} (or $r$-direction in a spherical coordinate) and the predicted Raman scattering broad wing profiles are not  affected by the height (see \citealt{cha15}).

%** Figure 4 **
\begin{figure}
  \centering
\includegraphics[width = 4.5 cm]{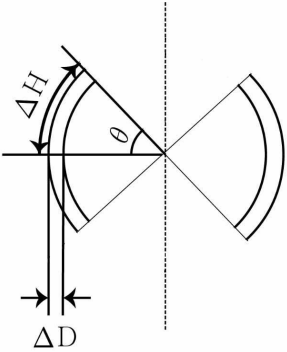}
\caption{A schematic diagram of the cross-section of a {\hi} accretion disk or a  {\hi} bipolar conic shell showing the thickness ($\triangle$D) and  height  (or arc length) ($\triangle$H). The vertical dotted line would correspond to the polar axis in case of an accretion disk geometry. }
\label{fig4}
\end{figure}

If the Ly$\beta$ optical depth meets the escape condition, $\tau_{Ly\beta,i(r)}$
$\geq$ $\tau_{r}$, the untransformed Ly$\beta$ will leave the {\hi} zone, having no chance to become the optical photon. Here, the $r$-component of  Ly$\beta$,  $\tau_{Ly\beta,i(r)}$ can be computed from the accumulated  $\tau_{Ly\beta}$ value at each scattering position. The computer program counts the entire number of the transformed optical photons in each wavelength segment from the above simulation procedure and constructs the Raman scattering line profile in the optical wavelength. Similar Monte Carlo simulation is to be done for Ly$\gamma$ photons as well in the same manner.
A similar procedure can be applied to the prediction of Raman scattering line profiles in the bipolar conic shell geometry.

\section{SYNTHETIC LINE PROFILES}

Due to the small scattering cross section, {\ie} $\sigma_{tot} \sim 10^{-22}$~cm$^2$, the Raman scattering mechanism requires the scattering region of a high column density (characterized by a thickness $\triangle$D $\times$ gas number density $n_{\rm H}$). For an active galactic nucleus (AGN) which engages the hard UV continuum, the large column densities, such as $N_{\rm H}$ = 10$^{22}$ -- $10^{24}$~cm$^{-2}$, corresponding to the total scattering optical depth, $\tau_{tot}$ = 1 -- 100 for $\lambda = $ 1011.83 -- 1035.69\AA\, around Ly$\beta$, are required to take in the Raman scattered photons effectively appear in the wavelength range of $\lambda = $ 6047 -- 7012\AA\, around {\ha}.
AG Peg does not demand such a high column density probably due to the relatively strong Ly$\beta$ and Ly$\gamma$ intensities.

Fig.~\ref{fig5} presents the plot of the synthetic {\hb} line profiles with  column densities of the shell thicknesses (much lower than those of the AGN cases), (a) 2 $\times 10^{20}$ cm$^{-2}$, (b) $10^{20}$ cm$^{-2}$, (c) 5 $\times 10^{19}$ cm$^{-2}$, and (d) 3 $\times 10^{19}$ cm$^{-2}$. The synthetic line profiles (dark solid line) were obtained by adding the recombination (dashed) and Raman scattering (solid) components. In the simulation, we assumed the blue and red Gaussian components to be at $-37$ {\kms} and +23 {\kms} (or at kinematic centers of $-30$ {\kms} and +30 {\kms}), adopting the observed {\hb} peak information in Table~\ref{tb3}. The continuum flux level, which is associated with the GS continuum and absorption line features, is not included in the prediction.
Note that the broad wing flux intensity increases with the column density while the nebular components remain to be the same: the higher the gas column density is, the stronger the broad wing flux becomes.

The simulation with the relatively high column densities of (a) 2 $\times 10^{20}$ cm$^{-2}$ or (b) $10^{20}$ cm$^{-2}$, predicted the Raman scattering component  stronger and broader than the observation (see the observed profiles in  Fig.~\ref{fig2}). We found that the column density simulation result with the relatively low column densities of (c) 5 $\times 10^{19}$ cm$^{-2}$ or (d) 3 $\times 10^{19}$ cm$^{-2}$ matches closely with the observed wing profile.

%** Figure 5 **

\begin{figure*}
\includegraphics[width = 11 cm]{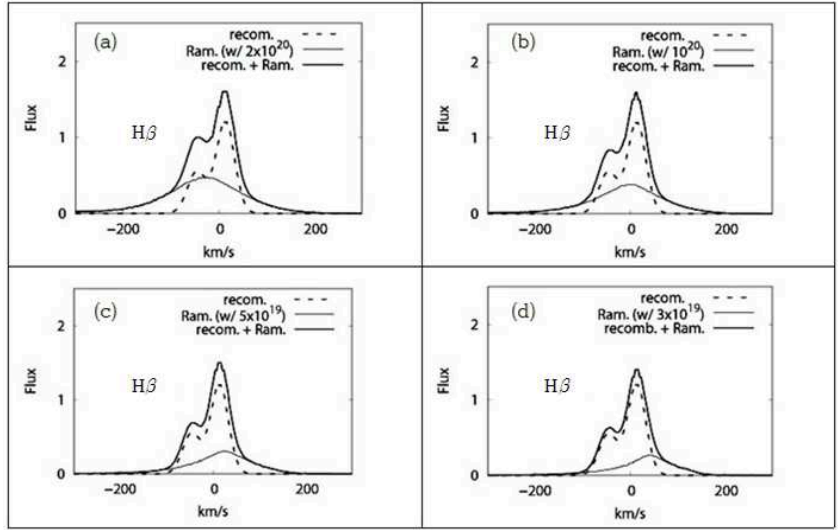}
\caption{Predicted {\hb} line profiles with FWHM = 65 {\kms} and with the blue and red peaks at $-$37 and 23 {\kms}. Dark solid: recombination + Raman scattering. Solid: Raman scattering. Dashed: recombination. Simulations with hydrogen column densities of the neutral hydrogen shell thickness: (a) with $N_{\rm H}$ = 2 $\times 10^{20}$ cm$^{-2}$ for the scattering neutral zone, (b) with $N_{\rm H}$ = $10^{20}$ cm$^{-2}$, (c) with $N_{\rm H}$ = 5 $\times 10^{19}$ cm$^{-2}$, and (d) with $N_{\rm H}$ = 3 $\times 10^{19}$ cm$^{-2}$.  Flux unit: 10$^{-13}$ erg s$^{-1}$ cm$^{-2}$ per {\kms} (same as in Fig.~\ref{fig2}). See the text. }
\label{fig5}
\end{figure*}

The prediction also shows that the red-shift placement of the central peak of the observed {\hb} broad wing profile is not an error. The predicted {\hb} broad wing profile  with the lower column density, $N_{\rm H}$ = 3 $\times 10^{19}$ cm$^{-2}$ in Fig.~\ref{fig5}, closely matches the observed profiles  for both the broad line center and width. The wavelength dependency of the total cross sections is asymmetric around Ly$\gamma$, having larger values in the blue part than in the red portion (see \citealt{cha15}: Fig.~1). A combination of both the asymmetric wavelength dependence of the cross-sections of scattering photons and the double Gaussian line profiles emitted in the {\hii} zone relocates the scattering line center at the reddish side.

Fig.~\ref{fig6} shows the predicted {\ha} and {\hb} line profiles with a slightly larger peak separation at -45 and 25 {\kms}. The top {\ha} and {\hb} profiles in Fig.~\ref{fig6} (a) and (b) show the predicted plots  with a {\hi} column density of $N_{\rm H}$ = 5 $\times 10^{19}$ cm$^{-2}$, while the bottom {\ha} and {\hb} profiles in Fig.~\ref{fig6} (c) and (d) are predicted profiles with $N_{\rm H}$ = 3 $\times 10^{19}$ cm$^{-2}$. The predicted {\ha} line with $N_{\rm H}$ = 5 $\times 10^{19}$ cm$^{-2}$ fits the observed profiles better, while the predicted {\hb} line  with $N_{\rm H}$ = 3 $\times 10^{19}$ cm$^{-2}$ agrees with observation well, as seen in  Fig.~\ref{fig6} (a) and (d).

%** Figure 6 **

\begin{figure*}
\includegraphics[width = 11 cm]{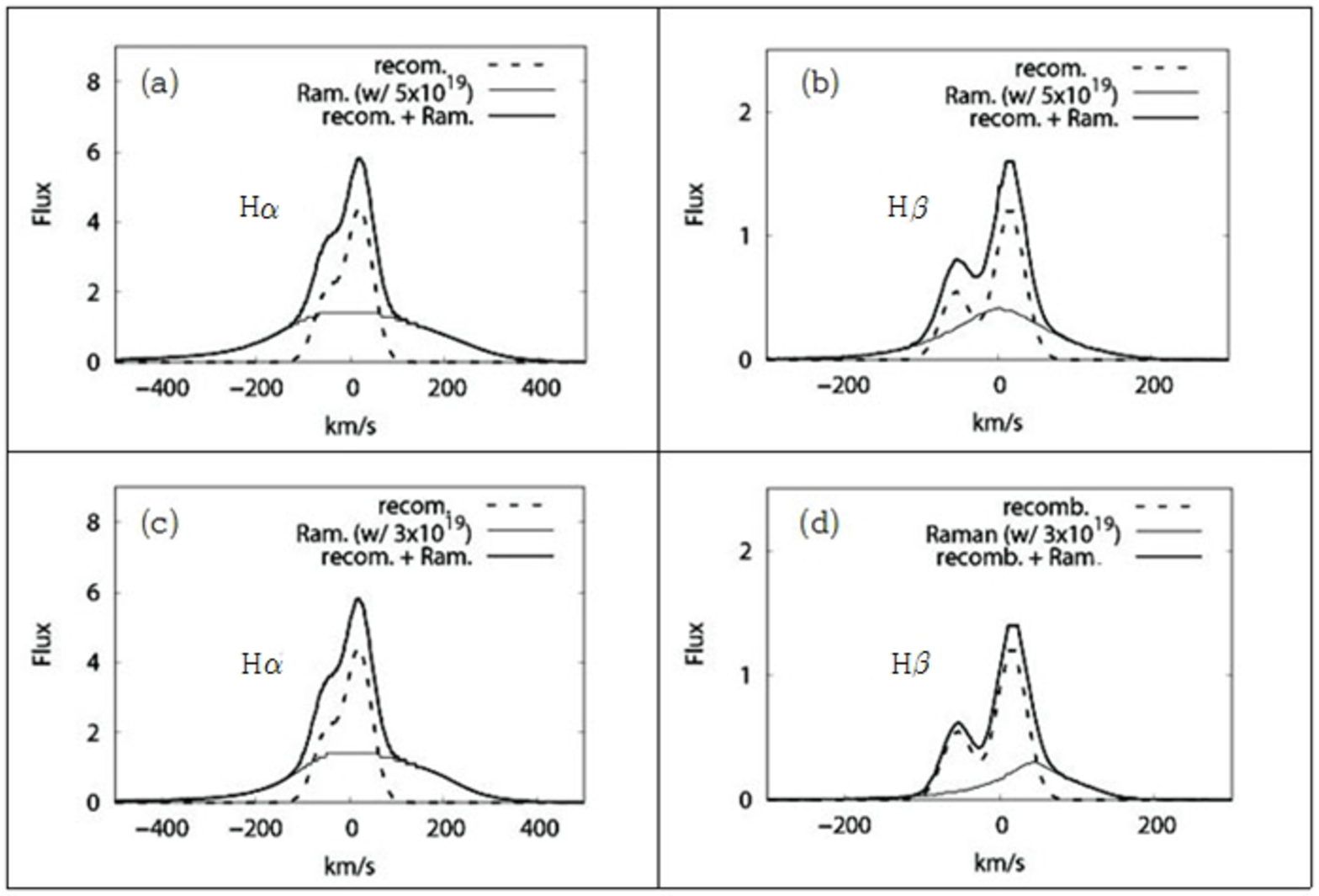}
\caption{Predicted {\ha} and {\hb} line profiles with the FWHM = 65 {\kms} and with the blue and red peaks at $-$45 and 25 {\kms}, respectively. (a) and (b):  with $N_{\rm H}$ =  5 $\times 10^{19}$ cm$^{-2}$ for the neutral hydrogen zone.
(c) and (d): with $N_{\rm H}$  = 3 $\times 10^{19}$ cm$^{-2}$. See Fig.~\ref{fig5} and the text. }
\label{fig6}
\end{figure*}

Table~\ref{tb6} summarizes the fractions of the predicted values of the Raman scattering and the recombination fluxes by the two models. The predicted nebular and Raman scattering portions of the {\hb} flux of $N_{\rm H}$ = 3 $\times 10^{19}$ cm$^{-2}$ are 71\% and 29\%, respectively, close to those of observation, 80\% and 20\%, respectively (in Table~1). The observed nebular and broad wing components of the {\ha} profile are  35\% and 65\%, respectively, while the predicted fractions with  $N_{\rm H}$ = 3 $\times 10^{19}$ cm$^{-2}$ are 46\% and 54\%, respectively, which differ with the observation largely. Meanwhile, the predicted values,  42\% and 58\% for the nebular and broad wing components with $N_{\rm H}$ = 5 $\times 10^{19}$ cm$^{-2}$ appear to be close to the observation.
The Raman scattering simulation fitted the observed FWHMs  and relative flux of the broad lines fairly well.

%** table 6 **

\begin{table}
\caption{Nebular vs. Raman wing flux fraction.
See Table~1 for  the flux measurements. The predicted {\hb} line shows that  the column density of the {\hi} zone is about 3 $\times 10^{19}$ cm$^{-2}$, while  the {\ha} seems to indicate $N_{\rm H}$  =  5 $\times 10^{19}$ or higher.}
\vspace{-0.5cm}
\begin{tabular}{cccc}\\
\hline \hline

& observed & predicted & column density \\

\hline
lines          &  Nebular  : Raman     &  Nebular  : Raman      &  $N_{\rm H}$ (cm$^{-2}$)  \\
{\ha}           &   35\%  :  65\%      & 42\%  :   58\%          &  5 $\times 10^{19}$  \\
              &                          & 46\%  : 54\%             &  3 $\times 10^{19}$  \\
{\hb}           & 80\% : 20\%           &71\% :  29\%            &  3 $\times 10^{19}$  \\
\hline

\end{tabular} \\
\label{tb6}
\end{table}

\section{KINEMATICAL STRUCTURE FOR THE EMISSION ZONE}

So far, we concentrated on the Raman scattering line formation process, responsible for the broad wing component formed in the {\hi} shell and skipped any in-depth discussion on the kinematic properties of the {\hii} region responsible for the double Gaussian line profiles. We proved that the Ly$\beta$ and Ly$\gamma$ photons with a double Gaussian profile form would eventually metamorphose into the single Gaussian broad {\ha} and {\hb} line components, similar to the observed line profiles. Comparison of the Raman scattering  line profile with the pure recombination line profile would give a tip on the kinematic structure responsible for the observed emission and scattering zones.

The performed Monte Carlo simulation simply assumed that the inner emission {\hii} region located close to the WD, emitted hydrogen emission lines in double Gaussian line profile form.
The measured FWHMs for the three components of the {\ha} and {\hb} lines are given in  Table~\ref{tb3} (see Fig.~\ref{fig2}). After correcting the kinematic radial velocity, $V_{\rm r}$ = $-19.09\pm$0.14, the centers of the double Gaussian line profiles would be  at  $-$36.1 and +35.1 {\kms}, respectively, while the observed FWHM was about 56 -- 65 {\kms}. We adopted FWHM = 65 {\kms} and assumed the double Gaussian peak separation at  $-$35 and +35 {\kms} (or at $-$30 and +30 {\kms}) in the simulation to accommodate the observed broad line profiles.

We examined two different models given in Table~\ref{tb5}: (1) the rotating accretion disk geometry in Model I-B and Model II-A
and (2) the expanding bipolar outflows geometry in Model I-C and Model II-B.
If one wants to adopt one of the two model geometries as the origin of the double peaks, one must first examine its validity based on other relevant physical parameters responsible for the assumed double line profiles at  $-30$ {\kms} and +30 {\kms} or  $-35$ {\kms} and +35 {\kms}.
%whether the relatively large path at both edges of the accretion disk
%would produce the saddle type
%We will check its validity from the Keplerian rotation velocity.
Such physical parameters would be the Keplerian rotation velocity of the disk around the WD and the binary separation between the two stars in the case of the accretion disk model. Table 5 lists some check points to determine validity of the assumed model geometries to fit the observed double Gaussian lines in AG Peg.

The AG Peg's orbital period is 816.5 days and the GS is classified as M3 III \citep{sch88}. The mass of the hot WD in symbiotic stars is known to be larger than that of planetary nebulae. Hence, with the WD M$_{\rm WD}$ $\sim$ 0.65 -- 1.4 M$_{\odot}$ and the GS M$_{\rm GS}$ $\sim$ 2 -- 6 M$_{\odot}$, the binary separation {between the two stars} is in a range of 2.4 -- 3.3 AU.
\citet{form90} suggested the larger side value, \ie 3.3~AU. \citet{con97} derived  the WD radius R$_{\rm WD}$ = 0.064R$_\odot$ and the GS radius R$_{\rm GS}$ = 78 R$_\odot$ (or 0.37 AU) for d = 650 pc (see also \citealt{mur95}). The radius of the GS is relatively smaller than the binary separation and then the binary system would be a detached system (that might be different from the example of Fig. 3).

Assuming the WD mass M$_{\rm WD}$ $\sim$ 1.4 M$_{\odot}$, we derive the Keplerian speed for Model I-B,
$$
{V_{\rm H\,_{\sc II}}} = \sqrt{\frac{G{\rm M_{\rm WD}}}{{r_{\rm H\,_{\sc II}}}}} = 30 \sqrt{\frac{1.4}{2.1} } = 24.5~ {\rm km s^{-1}},
$$
which is smaller than the observed double peak separation at $-35$ and +35 {\kms}. The Model II-A accretion disk gives an even slower Keplerian speed, V$_{\rm HII}$  =  22.7~{\kms}. These two $\sim$2.1 -- 2.45 AU accretion disk models cannot accommodate the observed double peak separation at $-35$ and +35 {\kms}, rejecting the possibility of the accretion models as the geometrical structure(s) responsible for the observed emission lines.

Forthwith, we will examine the other alternative, the bipolar conic model in Model I-C and Model II-B in Table~5. Although the shell of AG Peg might be opaque along the equatorial zone, the relatively low density gas zones might have shaped along the bipolar directions due to the WD winds. Such relatively low density bipolar outflows from the WD would be ionized by the UV radiation from the WD. As a consequence, the inner region of the bipolar conic shells close to the WD is to be fully ionized and becomes the {\hii} zone responsible for the observed double Gaussian profiles.

{% while the right edge (receding from the observer) appears to be more voluminous than the left edge. }

%** Figure 7 **
\begin{figure}
  \centering
\includegraphics[width = 4.0 cm]{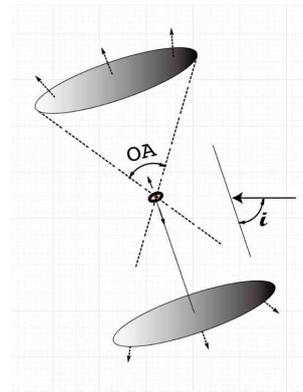}
\caption{A schematic diagram of a bipolar conic shell geometry. The model geometry assumes the same opening angle ($OA$) for both bipolar conic shells. The inclination angle ($i \sim 60^{\circ}$) of the pole is relative to the observer as specified on the right side. The arrows near the cones indicate the expansion of the bipolar conic shells. }
\label{fig7}
\end{figure}

Table~\ref{tb7} summarizes the parameters for the model conic shells and the predicted FWHMs, corresponding to Model I-C or Model II-B in Table~5.
The bipolar conic shells are assumed to expand at constant velocities, V$_{\rm exp}$  =  70 {\kms} and V$_{\rm exp}$ = 100 {\kms}, for Case A and Case B, respectively.
In order to accommodate the observed double Gaussian peaks at $-35$ and +35 {\kms}, the polar axis is taken to be slanted with an inclination angle \eg  $i =$ 60$^{\circ}$ or  $i =$ 70$^{\circ}$, respectively, relative to the line of sight in Case A and Case B.
Three different opening angles, $OA = 74^{\circ},~ 55^{\circ}$,~and~$35^{\circ}$, are considered to examine the kinematic line width.
Table~\ref{tb7} shows that the bipolar conic shell either $OA = 55^{\circ}$ in Case A or ~$35^{\circ}$  in Case B would be able to fit the observed FWHMs.

%** table 7 **

\begin{table}
\caption{Physical parameters and predicted FWHMs of bipolar conic shells.
The inclination angle ($i$) of the polar axis relative to the line of sight is adapted to fit the double Gaussian peaks at $\pm$35 {\kms}. V$_{\rm exp}$ and FWHM: {\kms}.
See Fig.~\ref{fig7} and Fig.~\ref{fig8}.}
\vspace{-0.5cm}
\begin{tabular}{lcc}\\
\hline \hline

parameter & Case A & Case B \\

\hline
V$_{\rm exp}$   &  70   & 100  \\
inclination angle ($i$) &   60$^\circ$ &  70$^\circ$ \\
%\hline
%{\bf $<$opening angle$>$ } = & {\bf $40^{\rm \circ}$ }   & {\bf  $40^{\circ}$ }  \\
%FWB (FWHM)  =   & 41.4 (38.9)   & 64.2 (60.3)   \\
\hline
{\bf $<$opening angle$>$  }  & {\bf $74^{\circ}$  } &  {\bf $74^{\circ}$ } \\
~~~FWHM   & 68.5    &  106   \\
\hline
{\bf $<$opening angle$>$  }  & {\bf $55^{\circ}$  }  &  {\bf $35^{\circ}$ } \\
~~~FWHM     &  52.5   &   53.1   \\
\hline
Obs'd FWHM   &  44 -- 55  & 44 -- 55  \\
\hline

\end{tabular} \\
\label{tb7}
\end{table}

Fig.~\ref{fig7} shows a schematic diagram for $OA = 74^{\circ}$ in Case A.
The line of sight is specified on the right side, for an inclination angle $i \sim 60^{\circ}$ of the polar axis. The small arrows show the expansion of the bipolar conic shells. Since the observer is assumed to be on the right side, the emission lines from the upper cone would appear to be red-shifted ($+35$ {\kms}), while those from the lower cone would be blue-shifted ($-35$ {\kms}).

Fig.~\ref{fig8} shows the synthetic line profiles from the upper cone (\ie receding relative to the observer) of Fig.~\ref{fig7}.
The synthetic line profile was constructed with the observed spectra indicated Gaussian smoothing factor $\sigma_G$ = 35 {\kms}.
We did not show the blue component line profile at $-$35 {\kms} which has a bilateral symmetric shape with a relatively weak strength.

%** Figure 8 **
\begin{figure}
  \centering
\includegraphics[width = 3.5 cm]{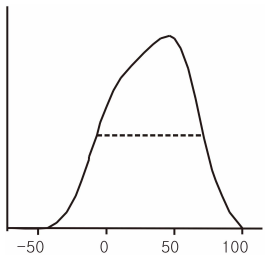}
\caption{Synthetic line profile for the receding component in Figure~\ref{fig7}.  Horizontal axis: {\kms}. Vertical axis: relative flux (not scaled).
The line profile is composed with a Gaussian smoothing factor  $\sigma_G$ = 35 {\kms}. Dashed line:  FWHM. }
\label{fig8}
\end{figure}

We demonstrated that a relatively simple single structure can accommodate both the narrow and the broad line components without taking for granted the presence of a colliding region between the GS and the WD \citep{nus95, mur95}. Moreover, we did not need to invoke a fast wind terminal velocity of 900 {\kms} \citep{pen85} to explain the broad line component.

The bipolar conic shells with faster expansion velocities, \ie Case B 100 {\kms} expansion model in Table~7, appear to be more appropriate than the low Case A 70 {\kms} expansion one. The bipolar conical outflows with the expansion velocity V$_{\rm exp}$ = 100 {\kms} with $OA$ $\geq$ 74$^{\circ}$ (Case B), can accommodate the much wider line profile, \ie FWHM $\sim$  114 {\kms}, corresponding to the {\ha} blue component in Table~3.

The main cause of the difference of the FWHMs seen in the observed double Gaussian profiles of Table~3 is likely to be the different opening angles of the bipolar cones, \eg 74$^{\circ}$ and 35$^{\circ}$ as presented in Table~7.
The small FWHM value of the red component in Table~\ref{tb3} corresponds to the small opening angle ($OA$ = 35$^{\circ}$), while the large FWHM of the blue component corresponds to the large opening angle ($OA$ $\simeq$ 74$^{\circ}$).

The scattering zone appears to be located in the outer part of the conical shell, perhaps attached to the radiation-bounded inner {\hii} zone, or slightly detached from it. The broad wing line profiles depend on both the physical circumstances of the inner ionization zone (involving the ionizing UV photons from the WD) and the column density of the thin outer neutral zone (\eg 0.035 times the thickness of the {\hii} shell).

If the {\hii} zone is radiation-bounded, the ionized zone stops so close to the physical edge of the material. If so, any slight change in the density of geometry of the outflow would cause the {\hi} zone along with the broad Balmer line components to disappear altogether.
Without the detailed analysis of the line profiles using the line analysis tool like Dipso, it would be hard to differentiate whether the broad line component exists in the observed line profiles by others.
Although we were not able to readily find any disappearance evidence of Raman scattering lines in AG Peg from the literature, we affirmed the significant change of the broad Balmer line components from the Lick observatory 2001 and 2002 HES data (Hyung and Lee, in preparation).

The adopted bipolar conic radius, 2.45 AU, in Model II-B, is slightly larger than the size of the GS upper atmosphere or the GS Roche lobe scale, $\sim$1.8 AU.
Depending on the inclination angle of the system, the GS upper atmosphere might block some fraction of the bipolar conic shell.  Monitoring the secular variation of the broad wing component might be critically important to look into the physical circumstance of the nebular gas, the accretion disk formation around the WD, and the mass loss activity from the GS.

Viewing the above limitations for the wind originates from the WD, we assumed that the proposed bipolar conic structure is actually part of the common envelope (CE) or the outer shell of the Roche lobe formed through the mass inflows from the giant star and pushed away by the fast winds from the hot WD. The swollen oval shape shell is introduced to maintain its thin shell structure due to the stable inflow from the GS despite of the orbital motion of the WD. The  high density compact accretion disk presumably present around the WD might prevent the UV radiation and the stellar winds along the equatorial zone reaching the thin CE, producing two bright parts of the CE, \ie similar to the bipolar conical shell appearance assumed in the work.

\section{CONCLUSIONS}

We successfully isolated  the broad wing component from the observed line profiles and investigated the Raman scattering process responsible for the  broad {\ha} and {\hb} line components through Monte Carlo simulations, assuming a bipolar conic shell formed from the WD winds. The physical conditions for the ionized {\hii} zone, \eg gas density, $n_{\rm H}$ = 10$^{9.85}$ cm$^{-3}$, and Balmer and Lyman line fluxes  and the WD luminosity, were taken from the P-I model and the column density for the scattering neutral zone was derived from the best prediction for broader lines.
{%  Isolating the broad component from the observed line profile was in fact a difficult job, in {\hb} line profile due to the adjacent {\niii} \& {heii} lines. However, the {\ha} line profile did not suffer such a problem. }

The radio observation by \citet{ken91} showed that the opening angles of the large scale bipolar contour seen in 1.5 and 5 GHz maps are very extensive (see 5 GHz contour image in their Fig. 5), while the present model investigation also indicates wide opening angles in the emission zones of the observed {\ha} and {\hb} lines. Such wide opening angles are likely to be structure-related to the inherent formation characteristics of the bipolar conic shell, which are actually the two highlighted parts of the swollen CE due to the UV and fast stellar winds from the WD. The large scale bipolar contour images seen in the radio maps might be the trace of the earlier small size bipolar conical shell formed around the WD.

Ignoring differences of the measured FWHMs between the blue and red Gaussian line profiles caused by several other unknown complexities, we carried out Monte Carlo simulations adopting the same FWHM, 65 {\kms} double Gaussian profiles (mean value) whose blue and red peak centers are at $-45$ and +25 {\kms} (or $-35$ and +35 {\kms}) for the hydrogen Balmer and Lyman synthetic line profiles. The main root of the line broadening appears to be the opening angle ($OA$) and the expanding outflows (V$_{\rm exp}$) of the bipolar conic shells.

We demonstrated that the expanding bipolar cones either with V$_{\rm exp}$ = $\pm$70 {\kms} and $i$ = 60$^{\circ}$ (Case A) or with V$_{\rm exp}$ = $\pm$100 {\kms} and $i$ = 70$^{\circ}$ (Case B), would be able to produce the double peak lines, at $-35$ and +35 {\kms}.
We also showed that either Case A model with $OA$ $\sim$ 55$^{\circ}$ or Case B model with  $OA$ $\sim$ 35$^{\circ}$ would be able to fit the observed FWHM ($\simeq$ 2$\triangle$V$_{\rm exp}$) = 44 -- 55 {\kms}. When bipolar polar winds are present in the system, double scattering can also take place, where intensity is transferred from the blue line to the red line due to an internal Doppler effect \citep{yoo02}.
We did not examine such a possibility.

{%The Gas dynamics responsible for the observed asymmetry of the double Gaussian profiles would be useful to verify the suggested conic structure in the future study. Further discussion is beyond the scope of the present study. }

The bipolar conic shells of r $\sim$ 2.5 AU, which might be two opposite regions of the swollen CE, is likely to be responsible for the  {\ha} and {\hb} double Gaussian line profiles. Adopting the ionization balance and structure suggested by the P-I investigation, we derive the thickness of the ionized {\hii} shell to be nearly $2.0 \times 10^{11}$ cm (in Model II-A: see Model II in Table~\ref{tb4}). Meanwhile the outer neutral part would be about 0.035 times thinner than the inner ionized zone, \ie 7.0 $\times 10^{9}$ cm for the column density, $N_{\rm H}$ = 5 $\times 10^{19}$ cm$^{-2}$.

In closing, we showed that the nebular double peak components and the broad Raman features were produced from the proposed bipolar conic zones, situated above and below the WD. The column density and other kinematical properties derived from the observed double Gaussian profiles do not seem to indicate that both {\hii} and {\hi} line zones are located in other positions, such as the GS upper atmospheres.  The bipolar cones appear to be the most probable structure responsible for the observed {\ha} and {\hb} double Gaussian and broad emission lines.

\section*{Acknowledgments}

We thank the anonymous referee for a careful review and many valuable suggestions. S.-J.L. and S.H. would like to acknowledge support from the Basic Science Research Program through the National Research Foundation of Korea (NRF 2015R1D1A3A01019370; NRF 2017R1D1A3B03029309).
We thank Chang, S.-J. and Lee, H.-W. for allowing us to use their Monte Carlo code. S.H. is grateful to the late Prof. Lawrence H. Aller of UCLA, who carried out the Hamilton Echelle observation program together with him.

\end{document}